Figure 1:

to the cross angular power *only from those objects common to both catalogues*. If it is assumed that in the most extreme case the *IRAS* galaxies are simply a subset of the optical galaxies then the shot noise could be $\simeq 500$. This has a small ($\lesssim 2\%$) effect on the final results when added to the expectation values. The other major approximation here is the assumption of a $4\pi$ sky coverage in the statistical comparison below. A full analysis would require the cross-talk of the incomplete sky harmonic coefficients to be taken in to account (Lahav, this volume). This is ignored here, consequently any errors are underestimated.

For a chosen power spectrum a maximum likelihood analysis can be performed for any number of harmonics to determine the best $b_I b_O \sigma^2_{8,matter}$. I have performed this analysis on the optical and *IRAS* catalogues described above for standard (shape parameter $\Gamma = 0.5$) and low density ($\Gamma = 0.2$) CDM models. Using harmonics $l = 1$ to $l = 10$ (to avoid non-linear scales) I find values of $b_I b_O \sigma^2_{8,matter} = 0.82 \pm 0.05$ for standard CDM and $b_I b_O \sigma^2_{8,matter} = 0.84 \pm 0.06$ for low density CDM. The joint normalisation is therefore remarkably indifferent to the assumed power spectrum (at least on these scales of 8 to $50h^{-1}Mpc$). ¿From other studies [3],[1],[4] the real space value of $b_I \sigma_{8,matter}$ has been determined as $\simeq 0.7$. It is therefore possible to make a simple estimate of $b_O/b_I$ since $b_O/bI = (b_I b_O \sigma^2_{8,matter})/(b_I \sigma_{8,matter})^2$. I therefore estimate $b_O/b_I \simeq 1.7$ on the scales of 8 to $50h^{-1}Mpc$ from this analysis. As a cautionary note, Figure 1 shows the harmonic reconstruction number density as a function of distance (in arbitrary units) along a line close to the SuperGalactic Plane (using harmonics to $l_{max} = 10$) for the optical (solid curve) and *IRAS* (dashed curve) galaxies. The reconstructions have been renormalised according to the mean density of optical and *IRAS* galaxies. Centaurus/Great Attractor and Perseus-Pisces are labelled, the vertical dotted lines are at $b = -90°, 0°$ and $90°$. Clearly, while both populations trace the same underlying structure there is a marked difference at the peaks (the differences in sample depths is also a factor). Work in progress will address the issue of relative bias as a function of scale, using the SHA approach.

# COSMOLOGICAL PARAMETERS FROM SPHERICAL HARMONIC ANALYSES

Caleb. A. Scharf
*Institute of Astronomy, Madingley Rd., Cambridge, CB3 0HA.*

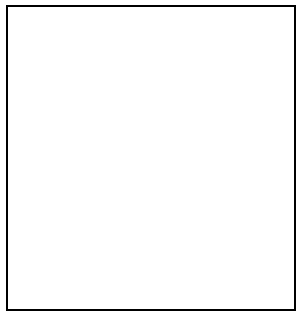

**Abstract**

In the rest of this volume Lahav and Fisher have presented the method of Spherical Harmonic Analysis (SHA) and its use in estimating cosmologically interesting parameters. Here I demonstrate the use of SHA in cross-correlating galaxy catalogues and present an estimate of the ratio of the optical and *IRAS* galaxy bias parameters; $b_O/b_I \simeq 1.7$.

## 1  METHOD AND RESULTS

The spherical harmonic transform of the distribution of objects on the full sky can be found as a sum over objects $a_{lm} = \sum_i f(r_i) Y^\star_{lm}(\hat{\mathbf{r}}_i)$, where $f$ is an arbitrary radial weight and $Y_{lm}$ are the orthonormal spherical harmonics. An estimator of the cross angular power between two catalogues ($I$ and $O$) can be written as:

$$c_{l,IO} = \frac{1}{(\Omega_{obs}/4\pi)(2l+1)}[Re(a_{l0,I})Re(a_{l0,O}) + 2\sum_{m=1}^{l}(Re(a_{lm,I})Re(a_{lm,O}) + Im(a_{lm,I})Im(a_{lm,O}))] . \tag{1}$$

The model prediction can be written as:

$$\langle a_{l,I}.a_{l,O}\rangle = b_I b_O \sigma^2_{8,matter} \frac{2}{\pi}\int k^2 P(k) \Psi^R_{l,I}(k)\Psi^R_{l,O}(k) dk , \tag{2}$$

where $b_I b_O$ is the product of the *IRAS* and optical bias parameters, $\sigma^2_{8,matter}$ is the rms variance in density in $8h^{-1}Mpc$ spheres, $P(k)$ is the unnormalised power spectrum of matter fluctuations and $\Psi^R_{l,I}$ and $\Psi^R_{l,O}$ are the appropriate harmonic window functions for *IRAS* and optical galaxies in real space, as described by Lahav (this volume). Here a simple number weighted ($f(r) = 1$) harmonic cross-correlation is applied to the *2Jy IRAS* catalogue [5] and a merged UGC and ESO optical catalogue constructed by M. Hudson [2]. The optical catalogue has a sky coverage of $\simeq 67\%$ (12177 objects), here I have used a *2Jy* catalogue with unsurveyed regions filled according to [6] and have then imposed the same limited coverage as in the optical (leaving 2137 objects). There will be a shot noise contribution